\documentclass[amsmath,amssymb,aps,twocolumn]{revtex4}
\makeatletter 
\def\@dotsep{4.5}
\makeatother

\usepackage{graphicx}
\usepackage{bm}

\begin{document}


\title{A High Speed, Post-Processing Free, Quantum Random Number Generator}

\author{J. F. Dynes}
 \email{james.dynes@crl.toshiba.co.uk}
\author{Z. L. Yuan}
 \email{zhiliang.yuan@crl.toshiba.co.uk}
\author{A. W. Sharpe}
\author{A. J. Shields}
\affiliation{Toshiba Research Europe Limited, Cambridge Research Laboratory, 208 Cambridge Science Park, Milton Road, Cambridge, CB4 0GZ, UK}

\date{\today}
\begin{abstract}
A quantum random number generator  (QRNG) based on gated single photon detection of an InGaAs photodiode at GHz frequency is demonstrated. Owing to the extremely long coherence time of each photon, each photons' wavefuntion extends over many gating cycles of the photodiode. The collapse of the photon wavefunction on random gating cycles as well as photon random arrival time detection events are used to generate sequences of random bits at a rate of $4.01$ megabits/s. Importantly, the random outputs are intrinsically bias-free and require no post-processing procedure to pass random number statistical tests,  making this QRNG an extremely simple device. 
\end{abstract}

\maketitle
Random number generators (RNGs) are important devices for many branches of science and technology. They find applications ranging from medicine \cite{tu2005} to testing fundamental principles of physics \cite{jacques2007}. Arguably the most important technological application of RNGs is cryptography \cite{schneier1996}. The requirements of the application usually determine the particular type of RNG chosen. Fast and easily obtainable random numbers can be generated from a simple computer algorithm, suitable for Monte Carlo simulations for example \cite{metropolis1949}. However, such an RNG is termed ``pseudo'' random as the random numbers are based on a deterministic pattern that will eventually repeat itself. The random number pattern is completely predictable. An RNG based on physical noise or some classical chaotic process improves upon the pseudo-RNG in terms of unpredictability although the bit rate of a physical RNG is usually much slower, limited to a few hundred kilohertz \cite{xu2006}.

Complete unpredictability of random number generation is desirable for a number of applications, notably  most demonstrations of quantum key distribution (QKD) \cite{gisin2007} which requires keys with high quality random bits.  A useful advance in realizing high quality RNGs has been to utilize quantum mechanical uncertainty in the generation of random numbers \cite{schmidt1970}, so-called quantum RNGs (QRNGs). Some of the most efficient and high bit rate QRNG demonstrations over the past few years have been based on single photon detection in conjunction with either spatial or temporal discrimination or a combination of both\cite{Jennewin2000,stipcevic2007}. Spatial discrimination is simply realized by use of a 50:50 beamsplitter with two single photon detectors; each detector monitors one of the beamsplitter output ports. A source of photons, conveniently derived from a laser attenuated down to the single photon level, impinges upon the input port of the beamsplitter. There is a $\frac{1}{2}$ probability that one of the beamsplitter output port detectors will measure an incident photon. Thus a random binary digit bitstream can be realized based on quantum mechanical uncertainty. Unfortunately from a practical point of view the random bits generated will not be completely free from bias as the two detectors will almost certainly have unequal detection efficiencies and it is very difficult to ensure the beamsplitter will have exactly 50:50 transmission reflection probabilty. The first problem can be somewhat circumvented by using one detector \cite{stefanov2000}. However, the second problem is very difficult to solve. Therefore, post bit collection processing is inevitable to unbias the raw random bits. 

\begin{figure}[!pb]
\centerline{\scalebox{1}{\rotatebox{0}{\includegraphics*[0mm,0mm][75mm,60mm]{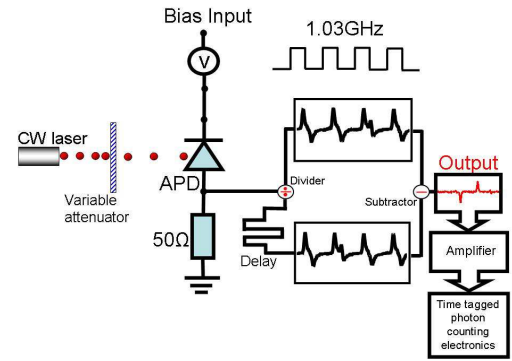}}}}
\caption{\label{fig:experimental_setup} (Color online) Schematic of the QRNG setup. A DFB CW laser (1550nm) attenuated approximately to the single photon level impinges upon an InGaAs APD. The APD is biased at a DC level of 45.9V and a superimposed a square wave modulation of amplitude 6V and frequency 1.03GHz which drives the device above breakdown. The resulting signal is sent through a self-differencing circuit, to remove the capacitive response, before being amplified and then finally sent to time tagging single photon counting electronics.}
\end{figure}
 
We note a useful QRNG based on temporal discrimnation of photon arrival times has been demonstrated recently \cite{stipcevic2007} achieving a random bit rate of $\sim1$Mbit/s. However, similar to many of the spatial discrimnation schemes, some post processing of the raw bits was needed to produce random bits of high enough quality.

In this paper we propose and demonstrate a new technique to generate random bits quantumly. It uses a fundamentally different method to those QRNGs proposed previously. 
It is a simple device requiring only one source of photons (a CW laser attenuated to the single photon level) and one single photon detector. The photons are emitted at random intervals determined by the quantum mechanics of photonic emission. They also have very long ($\sim 1\mu$s) coherence times. This long photon coherence time means the photon wavefunction extends over many gating cycles of the detetor. The collapse of the photon wavefunction on a random detector gating cycle signals a detection event and these random arrival events are converted into sequences of random bits. Furthermore, the QRNG satisfies two key requirements: it outputs high quality randomness and at a high rate without any need for classical post-processing either in software or hardware.


The single photon detector we employ is an InGaAs avalanche photodiode (APD), operating in gated Geiger mode. As is well known, InGaAs APD's can only be operated up to a few megahertz due to the unavoidable onset of afterpulsing \cite{ribordy1998}. However, a technique of electronic self-differencing shown previously \cite{yuan2007b}, can allow the detector to measure extremely small avalanches permitting considerably smaller bias voltages and hence much reduced after-pulsing. Consequently, the detector can be driven at very high frequency ($>1$GHz) before after-pulsing becomes problematic. 

\begin{figure}[!pb]
\centerline{\scalebox{0.7}{\rotatebox{0}{\includegraphics*[0mm,6mm][108mm,80mm]{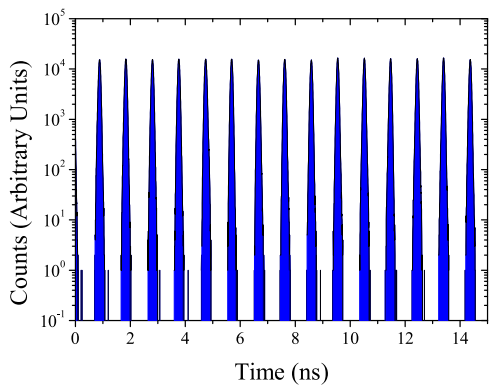}}}}
\caption{\label{fig:histogram} (Color online) A histogram of the photon detections in individual clock cycles from the QRNG setup under CW illumination.}
\end{figure}

\begin{figure}[!pb]
\centerline{\scalebox{1}{\rotatebox{0}{\includegraphics*[0mm,0mm][80mm,120mm]{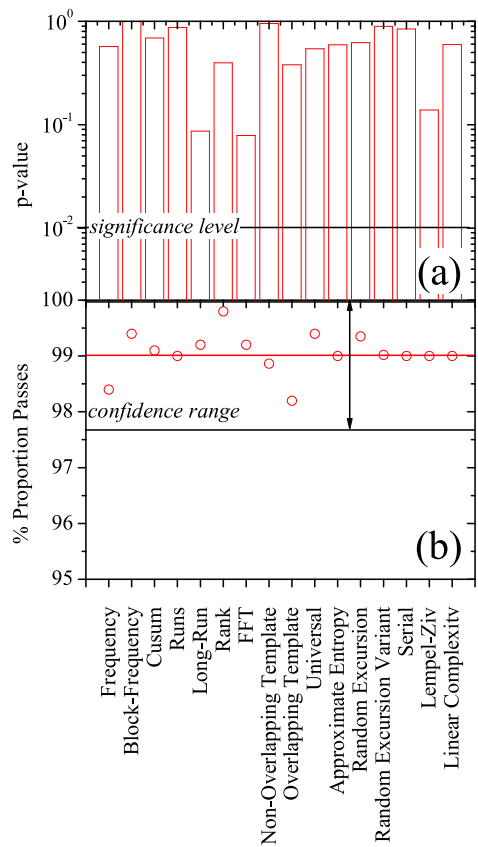}}}}
\caption{\label{fig:nist}(Color online) (a) Results of the NIST statistical tests on $500\times10^{6}$ bit patterns of binary bits from the quantum-RNG. All tests are passed at the $\alpha=0.01$ significance level. (b) The proportion of passes of each NIST test for the 500 bit streams.}
\end{figure}
The operation of the device is indicated schematically in figure \ref{fig:experimental_setup}. A coherent CW 1550nm laser diode, attenuated to the picowatt level, is coupled using a single mode fiber directly into an InGaAs APD. The APD is held at a temperature of $-30^{o}$C. Under normal operating conditions the large capacitive response of the APD obscures weak photon induced avalanches. A GHz square wave bias pulse train of approximately 6V amplitude is superimposed upon a DC bias of 45.9V, which is approximately 1.5V below its breakdown value. The APD signal is divided into two arms and one arm is delayed by one clock cycle before the two arms are recombined using a differencer. As can be seen, the large capacitive repsonse of the APD is cancelled leaving behind the weak avalanche signal\cite{yuan2007b}. With this setup, the InGaAs dark count rate is $\sim 10$kHz. The remaining signal is amplified before being sent into time tagging, single photon counting electronics rated up to 5MHz.

Figure \ref{fig:histogram} shows a histogram of the counts recorded for a detector clock frequency of 1.03GHz and a photon incident flux of approximately 20pW. A peak can be seen in each clock cycle completely separated from each other in time by $\sim 600$ps where no counts are recorded. Any overlap of adjacent clock counts could lead to unwanted correlations in the final random bit output. The sharp and narrow measured width  ($\sim 60 - 70$ps) of the clock cycle events is due mainly to the avalanche occuring over a small interval of the 500ps clock bias period. 

\begin{figure}[!pb]
\centerline{\scalebox{1}{\rotatebox{0}{\includegraphics*[5mm,5mm][85mm,120mm]{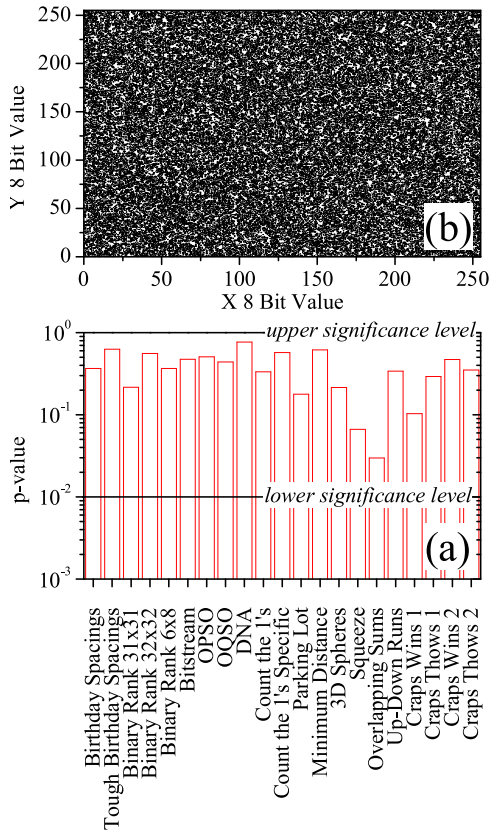}}}}
\caption{\label{fig:diehard}(Color online) (a) Results of the DIEHARD statistical tests on a block of $5\times10^{8}$ bit pattern of binary bits from the QRNG. All tests are passed at the $\alpha=0.01 \rightarrow \alpha=0.99$ significance level. (b) Byte correlation plot of the $5\times10^{8}$ bit pattern.}
\end{figure}
 
\begin{figure}[!pb]
\centerline{\scalebox{1}{\rotatebox{0}{\includegraphics*[15mm,0mm][75mm,60mm]{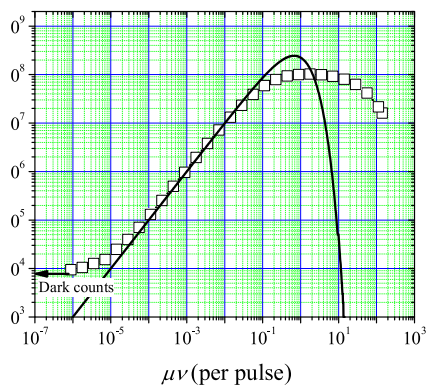}}}}
\caption{\label{fig:detector_rates}(Color online) Calculation depicting the self-differencing count rate of the detector (solid line) and  the experimentally measured self-differencing output (squares) as a function of the product of photon flux and detector efficiency $\mu\nu$ for a clock frequency of $\sim 1$GHz. }
\end{figure}

Time tagged count events were recorded for a duration of $\sim 2$ minutes with a CW photon flux of approximately $20$pW at a clock frequency of $1.03$GHz. The photon flux corresponds to about $0.3$ photons per clock period. The time tagged events acquired were converted into random bits by assigning detection events in even clock cycles as a ``1" and detection events occcuring in odd clock cycles as a ``0". We acquired a large binary file of around 500 MBit recorded at a rate of $4.01$MBps. 

The $500$MBit random bit binary file was then subjected to two statistical test routines (1) the NIST statistical test suite \cite{nist1999} and (2) the Diehard test suite \cite{diehard1990}. The NIST test contains 16 individual tests and each test evaluates to a probability value or ``p-value". The p-value is defined as the probability a perfect RNG would generate the particular experimental test result or a result indicating a less random bit sequence \cite{nist1999}. A significance level $\alpha$ is also assigned for the test. Usually, for cryptographic applications, a significance level of $\alpha=0.01$ is chosen \cite{nist1999}. This means if the p-value is $>0.01$ the bit stream passes the test. Additionally because of statistical fluctuations, we expect 99\% of the bit patterns to pass a paticular test and 1\% to fail. The large experimental file was divided into 500 $\times 1$MBit separate smaller files, and these smaller files were statistically tested; the results are shown in Fig. \ref{fig:nist}(a). All tests are passed at the $\alpha=0.01$ significance level. Fig. \ref{fig:nist}(b) shows the proportion of passes for each test. The proportions are nicely distributed around the expected value of 99\%. A confidence range based on the finite number of the bitstreams tested (500) is also shown. All test results are within this confidence range of $97.6\rightarrow100$\% for the proportion of passes. For the second battery of statistical tests (DIEHARD) also reports p-value but the significance level is slightly differently defined. In DIEHARD's case, the siginificance level is rather a range, $\alpha = 0.01 \rightarrow 0.99$. Tests with p-values falling within this range are deemed to have passed the tests. For the block of $5\times10^{8}$ bits, 21 tests were applicable. Figure \ref{fig:diehard}(a) shows that all 21 tests are passed. Figure 4(b) shows an 8 bit correlation plot for the $5\times10^{8}$ bit pattern indicating visually the good quality of randomness for the QRNG.\cite{footnote}. 
 

Finally, we comment on the performance of the QRNG. Currently, the QRNG is limited by the time tagging electronics bandwidth which is at the moment $5$MHz. However, in principle the QRNG should be able to work up to around 100MHz. Figure \ref{fig:detector_rates} plots the experimental self-differencing output of the detector up to the saturation point at $100$MHz as a function of the incident photon flux $\mu$ and detector efficiency $\nu$ product: $\mu\nu$ (squares). Also plotted is the theoretical self-differencing output (solid line) which peaks at around $250$MHz. Both experimental data and theory drop off after their respective peaks as $\mu\nu$ further increases. This is due to the probability of photon events occuring in adjacent clock cycles becoming appreciable. Hence the self-differencing method cancels these adjacent events out leading to a reduction in the overall count rate. However, experimental data and theory differ at high $\mu\nu$ and this is due to our assumption in the theory that the amplitude of detector avalanches are photon number independent. Nevertheless, the experimental peak value of $100$MHz is the limit of the QRNG which is over two orders of magnitude higher than any bit rates of existing QRNG's. With improved time tagging single photon electronics and detector multiplexing a GHz QRNG could potentially be realised. 

In conclusion, we have demonstrated an extremely simple, QRNG which ouptuts random numbers at a rate of 4.01MHz. The random numbers have been subjected to stringent statistical tests and are found to pass all. The QRNG requires no post-processing of data and the bit rates can be increased further with higher bandwidth time recorded single photon counting equipment and/or through multiplexing. 

\newpage

\begin{thebibliography}{99}
\bibitem{footnote} We note that it was not possible to generate random data that passed all relevant statistical tests when the laser source was replaced with a broadband Light Emitting Diode (LED) emitting at 1550nm for the same bit rate. This illustrates the importance of long photon coherence times in producing high quality random numbers at high bit rates.
\end{thebibliography}
\end{document}